\documentclass[conference]{IEEEtran}
\IEEEoverridecommandlockouts
\usepackage{cite}
\usepackage{amsmath,amssymb,amsfonts}
\usepackage{algorithmic}
\usepackage{graphicx}
\usepackage{textcomp}
\usepackage{xcolor}
\usepackage{flushend,url}
\def\BibTeX{{\rm B\kern-.05em{\sc i\kern-.025em b}\kern-.08em
    T\kern-.1667em\lower.7ex\hbox{E}\kern-.125emX}}

\newtheorem{lemma}{Lemma}

\allowdisplaybreaks
\begin{document}

\title{STAR-RIS-aided RSMA for the URLLC\\ multi-user MIMO Downlink}

\author{
   \IEEEauthorblockN{Mohammad Soleymani$^1$, Ignacio Santamaria$^2$, Eduard Jorswieck$^3$, Robert Schober$^4$, Lajos Hanzo$^5$}\vspace{.2cm}
   \IEEEauthorblockA{
   $^1$University of Paderborn, Germany,  $^2$Universidad de Cantabria, Spain,
   $^3$ Technische Universit\"at Braunschweig, Germany\\
   $^4$Friedrich Alexander University of Erlangen-Nuremberg, Germany, 
   $^5$University of Southampton, United Kingdom\vspace{.2cm}
   \\
                     Email: \small{\protect\url{mohammad.soleymani@upb.de}}, \small{\protect\url{santamal@unican.es}}, \small{\protect\url{e.jorswieck@tu-bs.de}},
                     \small{\protect\url{robert.schober@fau.de}}, \small{\protect\url{lh@ecs.soton.ac.uk}}
}
}
\maketitle

\begin{abstract}
Rate splitting multiple access (RSMA) is intrinsically amalgamated with simultaneously transmitting and reflecting (STAR) reconfigurable intelligent surfaces (RIS) to enhance energy efficiency (EE) of the finite block length  (FBL) multiple-input multiple-output (MIMO) downlink. An alternating optimization-based algorithm is proposed to jointly optimize the transmit beamforming matrices, STAR-RIS configurations, and rate-splitting parameters. STAR-RIS attains 360-degree full-plane coverage, while RSMA provides a prominent gain by efficiently managing interference. Numerical results reveal a strong synergy between RSMA and STAR-RIS, demonstreating significant EE gains over reflective RIS and spatial division multiple access (SDMA).
\end{abstract}


\section{Introduction}

Next generation (NG) wireless communications will support a large number of devices with various services and improved sustainability in terms of energy and resource efficiency. The authors of \cite{wang2023road} provide a comprehensive overview of the visions, requirements, key technologies, and testbeds on the road to 6G, emphasizing the role of advanced network architectures and intelligent communication systems. Gong et al. in \cite{gong2022holographic} explore the theoretical foundations and enabling technologies of holographic multiple-input multiple-output (MIMO) communications, outlining future research directions for this promising 6G technology. Vaezi et al. in \cite{vaezi2022cellular} offer a detailed survey of advancements in 5G for cellular, wide-area, and non-terrestrial Internet of Things (IoT), outlining the path toward 6G networks. The paper in \cite{vaezi2022cellular} highlights the challenges and opportunities in integrating massive IoT connectivity, and low-latency communications for the NG Internet of Everything (IoE).

An attractive technology to face challenges of NG networks is the reconfigurable intelligent surface (RIS) \cite{wu2021intelligent, di2020smart}. Wu et al. \cite{wu2021intelligent} offer a comprehensive tutorial on intelligent reflecting surface (IRS)-aided wireless communications, detailing the system models, signal processing techniques and their performance analysis. In contrast, the authors of \cite{di2020smart} introduce the concept of smart radio environments empowered by RIS, discussing their operation, current research status and future challenges. Additionally, there are clear benefits of passive RIS in improving energy efficiency (EE) \cite{lyu2023energy, soleymani2024energy, soleymani2022rate, niu2023active, fotock2023energy,  soleymani2022noma}. Lyu et al. in \cite{lyu2023energy} explore hybrid RIS architectures in cell-free networks to enhance EE, while \cite{soleymani2024energy} compares different RIS setups in multiple-input single-output (MISO) broadcast channels. Our former work in \cite{soleymani2022rate} investigates integrating rate-splitting with RIS in MIMO systems under hardware impairments. Furthermore, \cite{niu2023active} analyzes the trade-off between spectral efficiency (SE) and EE in active RIS-assisted RSMA networks, and \cite{fotock2023energy} contrasts active and nearly passive RIS designs under global reflection constraints to optimize EE in RIS-aided systems.

For practical low-latency communications, finite blocklength (FBL) coding must be applied. The fundamental limits for ultra-reliable low-latency communications (URLLC) are established by analyzing channel coding rates within the FBL regime in \cite{polyanskiy2010channel}. By contrast, in \cite{erseghe2016coding}, tight performance bounds and their asymptotic approximations are derived using Laplace integral techniques for short-packet transmission scenarios. 

The benefits of RIS in enhancing the communication performance under FBL constraints, particularly for ultra-reliable low-latency communications (URLLC), are highlighted in \cite{soleymani2024rate2, li2021aerial, vu2022intelligent, soleymani2025energy, xie2021user, almekhlafi2021joint, soleymani2023spectral,   pala2023spectral,  soleymani2023optimization, katwe2024rsma, soleymani2024optimization, soleymani2024rate}. RISs are shown to significantly improve the SE versus EE trade-off in dynamic propagation environments. Several works, such as \cite{soleymani2024rate2}, \cite{vu2022intelligent}, and \cite{xie2021user}, investigate RIS-aided URLLC scenarios and demonstrate performance gains in short-packet transmissions through optimized beamforming and user grouping. Advanced architectures like simultaneously transmitting and reflecting (STAR)-RIS and aerial RISs are advocated in \cite{soleymani2023spectral} and \cite{katwe2024rsma}, respectively, for their ability to provide improved reliability and coverage. Moreover, the integration of rate-splitting multiple access (RSMA) with RIS, as examined in \cite{pala2023spectral} and \cite{soleymani2023optimization}, is shown to enhance robustness against channel impairments and improve throughput under stringent latency and reliability requirements. These studies established RIS as a key enabler for achieving efficient and reliable communication in the FBL regime.

As a future advance, the authors of \cite{mao2022rate} surveyed RSMA, detailing its theoretical foundations, its advantages in managing multi-user interference, and its superiority compared to conventional multiple access techniques such as spatial division multiple access (SDMA) and non-orthogonal multiple access (NOMA). The paper also surveys recent developments and practical applications across various network scenarios and outlines key future research directions, establishing RSMA as a promising paradigm for NG networks.

The RSMA approach can also be applied in low-latency communication situations using the FBL rate expressions that take coding length and error probabilities into account \cite{xu2023max}. The max-min fairness of RSMA in FBL communication scenarios is investigated in \cite{xu2023max}, showing that RSMA significantly enhances both the rate fairness and reliability in short-packet transmissions, making it well-suited for URLLC applications in NG wireless networks. 

In this paper, we demonstrate the advantages of STAR-RIS-aided RSMA in FBL scenarios, considering both fairness and EE. The concept of STAR-RIS was introduced in \cite{zhang2022intelligent, liu2021star} to improve the network performance for users on both sides of the RIS.  The existing literature, including \cite{ soleymani2024rate}, address this question only partly. It is shown in \cite{jorswieck2025urllc, soleymani2023noma2, soleymani2023energy} that the combination of STAR-RIS with user fairness for spatially separated user groups significantly improves performance. Furthermore, RSMA shows improvements in scenarios where the system is highly loaded and hence does not have sufficient degrees of freedom to mitigate interference. In  \cite{soleymani2024rate}, we proposed RSMA schemes to enhance SE and EE of MU-MIMO URLLC system, assisted by conventional reflective RIS. In this paper, we extend the results of \cite{soleymani2024rate} to STAR-RIS-assisted systems, revealing the beneficial synergy of RSMA and STAR-RIS. In particular, \textbf{our main contributions are}: 
\begin{itemize}
    \item We formulate the system model of a mode-switching-aided STAR-RIS-assisted MIMO broadcast channel (BC) and 1-layer rate splitting. The problem is to maximize the minimum individual EE under minimum rate constraints, sum transmit power constraints, and STAR-RIS mode-switching constraints. 
    
    \item We propose an alternating optimization (AO) based algorithm for  optimizing both the beamforming and STAR-RIS scattering matrices. The approach is based on the optimization framework proposed in \cite{soleymani2025framework} for fraction matrix programming problems, and we extend the solution proposed in \cite{soleymani2024rate} to STAR-RIS-aided systems. 
    
    \item Finally, we answer the question raised above by showing that reliable and energy-efficient network operation requires both STAR-RIS and RSMA for maximizing EE fairness. 
\end{itemize}

{\em Notations}: We represent scalar/vector/matrix by $x/{\bf x}/{\bf X}$. A zero-mean complex Gaussian vector ${\bf x}$ with covariance matrix ${\bf X}$ is denoted by ${\bf x}\sim\mathcal{CN}({\bf 0}, {\bf X} )$. Moreover, $Q^{-1}$, ${\bf I}$, $\text{Tr}({\bf X})$, $|{\bf X}|$, and $\mathfrak{R}(x)$ denote the inverse Q-function for Gaussian signals, an identity matrix, the trace of ${\bf X}$, the determinant of ${\bf X}$, and the real part of $x$, respectively.

\section{System Model and Problem Statement}
\label{sec:smps}
\begin{figure}[t]
	\centering
	\includegraphics[width=.6\linewidth]{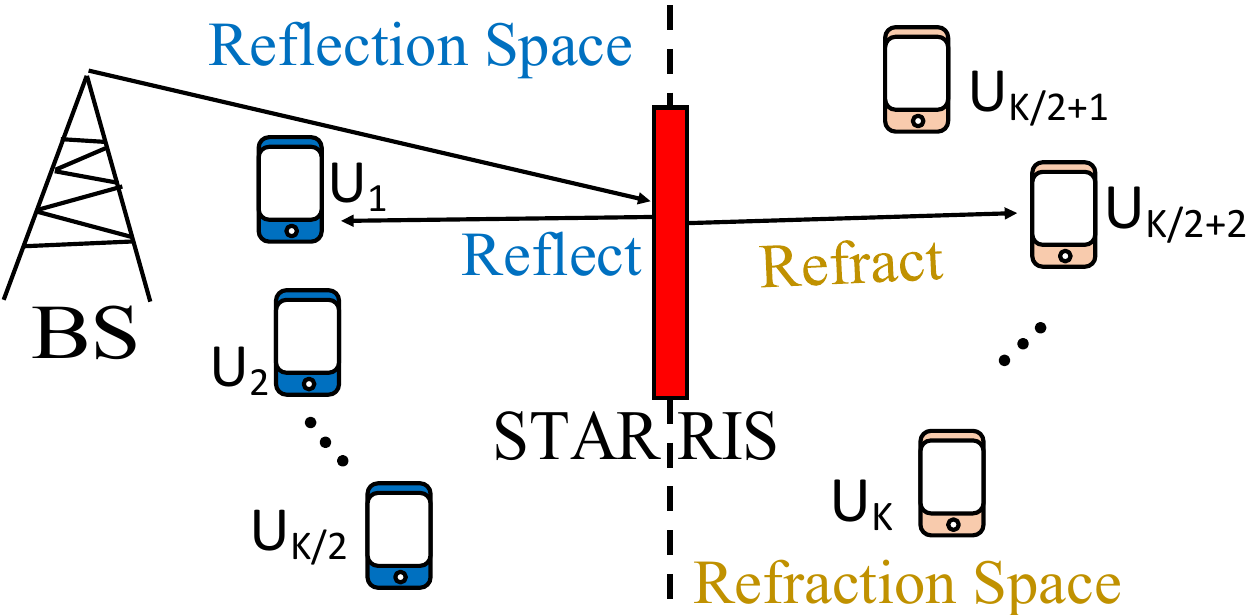}
	\caption{Network model: STAR-RIS-assisted downlink transmission.}\vspace{-.5cm}
	\label{fig:network-model}
\end{figure}

We consider a MIMO BC, where a base station (BS) equipped with $N_{BS}$ transmit antennas (TA) aims to support $K$ users, each equipped with $N_u$ receive antennas (RA). To enhance the communication performance, a STAR-RIS with $M$ elements is deployed to assist the BS, as shown in Fig. \ref{fig:network-model}.
Thus, the channel between the BS and user $k$ is modeled as  \cite{mu2021simultaneously}
\begin{eqnarray}
	\label{eq:cm}
	{\bf H}_k = {\bf D}_k {\bf \Theta}^{r/t} {\bf D} + {\bf G}_k, 
\end{eqnarray}
where ${\bf G}_k$ represents the channels between the BS and user $k$, ${\bf D}_k$ models the channels between user $k$ and the STAR-RIS, and ${\bf D}$ is the channel between the BS and the STAR-RIS. The matrices ${\bf \Theta}^{r} = \mathrm{diag}(\theta_1^r, \theta_2^r, \ldots, \theta_M^r)$ and ${\bf \Theta}^{t} = \mathrm{diag}(\theta_1^t, \theta_2^t, \ldots, \theta_M^t)$ represent the scattering matrices of the STAR-RIS corresponding to the reflection and transmission (refraction) modes, respectively. If a user is located in the reflection (or transmission) half-space, its channel can only be optimized through ${\bf \Theta}^{r}$ (or ${\bf \Theta}^{t}$). We assume a nearly passive RIS architecture, where the amplitude constraints of the RIS elements have to satisfy
    $|\theta^r_m|^2 + |\theta^t_m|^2 \leq 1$, for all $m$.
We consider a mode-switching scheme for STAR-RIS, as it improves the performance with lower complexities than the energy-splitting scheme. More specifically, we divide the STAR-RIS elements into two groups with an equal number of elements per group. The elements of the first group operate only in a transmission mode, while the elements of the second group operate in a reflection mode, i.e.,
\begin{subequations}\label{eq:con3}
\begin{align}
    \theta^r_m&=0, & |\theta^t_m|^2 &\leq 1, && \text{if} & 1&\leq m\leq \frac{M}{2},\\
     \theta^t_m&=0, & |\theta^r_m|^2 &\leq 1, && \text{if} & \frac{M}{2}+1&\leq m\leq M.
\end{align}
\end{subequations}

\subsection{Signal Model}

The BS leverages a 1-layer rate-splitting (RS) design, where the message of each user is divided into two parts: common and private. The common message is intended to be detected by all users, whereas the private messages are detected only by their respective intended users. More specifically, the signal transmitted from the BS is given by
	${\bf x} = {\bf W}_c {\bf t}_c + \sum_{k=1}^K {\bf W}_k {\bf t}_k$, 
where \( {\bf t}_c \sim \mathcal{CN}({\bf 0}, {\bf I}) \) denotes the common data stream, \( {\bf W}_c \) is its beamforming matrix, \( {\bf t}_k \sim \mathcal{CN}({\bf 0}, {\bf I}) \) is the private data stream intended for user \( k \), and \( {\bf W}_k \) is its corresponding beamforming matrix. Note that the common message aggregates the common parts of all users data. Moreover, we assume that the BS employs FBL coding for all transmitted data streams. The codeword lengths for the common and private messages are $n_c$ and $n_p$, respectively. 

Upon transmission of \( {\bf x} \), the signal  received by user \( k \) is ${\bf y}_k = {\bf H}_k {\bf W}_c {\bf t}_c + \sum_{j=1}^K {\bf H}_k {\bf W}_j {\bf t}_j + {\bf n}_k$,
where \( {\bf n}_k \sim \mathcal{CN}({\bf 0}, \sigma^2 {\bf I}) \) is the additive white Gaussian noise. 
User $k$ first detects ${\bf t}_c$, treating all other signals as noise, which results in the maximum achievable rate of \cite{polyanskiy2010channel, soleymani2024optimization}
\begin{multline}
    	r_{kc} =  \log\!\left| {\bf I} + \left(\sigma^2{\bf I}+\sum_{i} {\bf \Gamma}_{ki} \right)^{-1}{\bf \Gamma}_{kc} \right|\\
        -\frac{Q^{-1}(\epsilon_c)}{\sqrt{n_c}}  \sqrt{ 2\text{Tr}\left
    ({\bf \Gamma}_{kc}    \left[\sigma^2{\bf I}+ {\bf \Gamma}_{kc}+\sum_{i} {\bf \Gamma}_{ki}\right]^{-1} \right
    )}, \label{eq:cr}
\end{multline}
where ${\bf \Gamma}_{ki}={\bf H}_k{\bf W}_i{\bf W}_i^H{\bf H}_k^H$, ${\bf \Gamma}_{kc}={\bf H}_k{\bf W}_c{\bf W}_c^H{\bf H}_k^H$, and $\epsilon_c$ is the maximum tolerable detection error rate for ${\bf t}_c$. 
All users should be able to detect the common message, which is only possible when the transmission rate of ${\bf t}_c$ satisfies 
   $r_c \leq \min_k r_{kc}$.

After detecting ${\bf t}_c$, user $k$ removes it from ${\bf y}_k$ and then detects ${\bf t}_k$, treating all the remaining signals as noise. Therefore, the maximum achievable rate of ${\bf t}_k$ is 
\begin{multline}
    	r_{kp} =  \log\left| {\bf I} + \left(\sigma^2{\bf I}+\sum_{i\neq k} {\bf \Gamma}_{ki} \right)^{-1} {\bf \Gamma}_{kk} \right|\\
        - \frac{Q^{-1}(\epsilon_p)}{\sqrt{n_p}}
         \sqrt{ 
        {2\text{Tr}\left
    ({\bf \Gamma}_{kk}
    \left[\sigma^2{\bf I}+
    \sum_{i} {\bf \Gamma}_{ki}
    \right]^{-1} \right
    )}
    }, 
    \label{eq:cr}
\end{multline}
where  $\epsilon_p$ is the maximum tolerable detection error rate for ${\bf t}_k$. Note that the total detection error rate of each user can be upper bounded as $\epsilon_c+\epsilon_p$, as shown in \cite[Eq. (18)]{soleymani2023optimization}.

The rate of user $k$ is the summation of $r_{kp}$ and the portion of the rate it receives from detecting the common message, i.e.,
    $r_k=r_{kp}+q_k$,
where $\sum_kq_k=r_c$. 
Finally, the EE of user $k$ is \cite{buzzi2016survey}
\begin{equation}
    e_k=\frac{r_k}{p_k({\bf W}_c,{\bf W}_k)},
\end{equation}
where $p_k(\{{\bf W}_c,{\bf W}_k\})$ is the power consumed to transmit data to user $k$, given by \cite{soleymani2024rate}
\begin{equation}
    p_k({\bf W}_c,{\bf W}_k)\!=\!P_C\!+\!\beta(\text{Tr}({\bf W}_c{\bf W}_c^H )/K\!+\!\text{Tr}({\bf W}_k{\bf W}_k^H )),\!\!
\end{equation}
where $\beta$ is the power efficiency of the transmitter at the BS, and $P_C$ is the static power consumed to operate the network divided by $K$, given by \cite{soleymani2022improper}.

\subsection{Problem Statement}
We aim for maximizing the minimum EE of users, which can be formulated as 
\begin{subequations}\label{opt}
    \begin{align}    
	\max\limits_{{\bf W}_c,\{{\bf W}_k\}_{\forall k}, {\bf q}, {\bf \Theta}^r, {\bf \Theta}^t} & \, \min_{k } \frac{r_{pk}+q_k}{p_k({\bf W}_c,{\bf W}_k)} \label{eq:opt} \\
	\textrm{s.t.}\hspace{.5cm} & r_{pk}+q_k\geq \bar{r}_k^{th},\label{eq:con1}\\
    & \sum_kq_k\geq \min_k(r_{kc}),\label{eq:con4}\\
    &\text{Tr}\!\left(\!{\bf W}_c{\bf W}_c^H \! +\!\! \sum_{i=1}^K {\bf W}_i{\bf W}_i^H\! \right)\! \leq\! P,\! \label{eq:con2} \\
	&\eqref{eq:cm},\,\, \text{and}\,\,\, \eqref{eq:con3},\label{eq:con20}
\end{align}
\end{subequations}
where $P$ is the BS power budget, and $\{{\bf q} \}=\{q_1,q_2,\cdots,q_K\}$. The optimization problem in \eqref{opt} is non-convex. The objective function of \eqref{opt} is a fractional function, and its optimization variables include matrices. Therefore, \eqref{opt} is a fractional matrix programming problem. The constraints in \eqref{eq:con2} and \eqref{eq:con20} are convex. However, \eqref{eq:con1} and \eqref{eq:con4} are non-convex, since $r_{pk}$ and $r_{ck}$ are non-concave for all $k$.

\section{Proposed Solution}
\label{sec:aora}
 We solve \eqref{opt} by harnessing an iterative algorithm based on alternating optimization (AO) and the fractional matrix programming (FMP) solver proposed in \cite{soleymani2025framework}. We commence from a feasible initial point ${\bf W}_c^{(1)}$, $\left\{{\bf W}_k^{(1)} \right\}_{\forall k}$, ${\bf \Theta}^{r^{(1)}}$ and ${\bf \Theta}^{t^{(1)}}$. In each iteration, we first optimize the beamforming matrices, when ${\bf \Theta}^r$ and ${\bf \Theta}^t$ are kept fixed. We then alternate the optimization variables and update ${\bf \Theta}^r$ and ${\bf \Theta}^t$, while keeping the beamforming matrices fixed. In the following subsections, we provide the solutions for updating the beamforming matrices, STAR-RIS coefficients, and rate splitting parameters.

\subsection{Beamforming Optimization}
In this subsection, we propose a new solution for updating the beamforming matrices, when the RIS coefficients are kept fixed in iteration $j$. In this case,  \eqref{opt} is simplified to the following FMP problem
\begin{subequations}\label{opt-w}
    \begin{align}    
	\max\limits_{{\bf W}_c,\{{\bf W}_k\}_{\forall k}, {\bf q}} & \, \min_{k } \frac{r_{pk}+q_k}{p_k({\bf W}_c,{\bf W}_k)} \\
	\textrm{s.t.}\hspace{.5cm} & \eqref{eq:con1},\eqref{eq:con4}, \eqref{eq:con2}.
\end{align}
\end{subequations}
To solve \eqref{opt-w}, we employ the FMP solver proposed in \cite{soleymani2025framework}. To this end, we first calculate concave lower bounds for $r_{pk}$ and $r_{ck}$ in the following lemma.
\begin{lemma}[\!\!\cite{soleymani2024rate}]\label{lem:1}
    The following inequalities hold for all feasible ${\bf W}_c$, $\{{\bf W}_k\}_{\forall k}$
\begin{multline}\nonumber
r_{pk}\!\geq\! \tilde{r}_{pk}^{(j)}\! =\! 2\mathfrak{R}\!\left\{\!\text{{Tr}}\!\left(\!
{\bf A}_{k}{\bf W}_{k}^H
\bar{\mathbf{H}}_{k}^H\!\right)\!\right\}
+\!2\!\sum_{i\neq k}\!\mathfrak{R}\!\left\{\!\text{{Tr}}\!\left(\!
{\bf A}_{ki}{\bf W}_{i}^H
\bar{\mathbf{H}}_{k}^H\!\right)\!\right\}
\\
+a_{k}-
\text{{Tr}}\left(
{\bf B}_{pk}\left(\sigma^2{\bf I}\!+\!\!\sum_i\!\bar{\bf H}_{k}{\bf W}_{i}{\bf W}_{i}^H\bar{\bf H}_{k}^H\right)
\right),
\end{multline}
\begin{multline*}
r_{ck}\geq \tilde{r}_{ck}^{(j)} = a_{ck}
\\
+2\mathfrak{R}\left\{\text{{Tr}}\left(
{\bf A}_{ck}{\bf W}^H_c
\bar{\mathbf{H}}_{k}^H\right)\right\}
+2\sum_{i}\mathfrak{R}\left\{\text{{Tr}}\left(
{\bf A}_{cki}{\bf W}_{i}^H
\bar{\mathbf{H}}_{k}^H\right)\right\}
\\
-
\text{{Tr}}\!\!\left(\!\!
{\bf B}_{ck}\!\!\left(\!\!\sigma^2{\bf I}\!+\bar{\bf H}_{k}{\bf W}_c{\bf W}^H_c\bar{\bf H}_{k}^H+\!\!\sum_i\!\bar{\bf H}_{k}{\bf W}_{i}{\bf W}_{i}^H\bar{\bf H}_{k}^H\!\!\right)\!\!
\right),\!\!
\end{multline*}
where $\bar{\bf H}_k={\bf H}_k\left({\bf \Theta}^{r^{(j)}},{\bf \Theta}^{t^{(j)}} \right)$, and we have:
\begin{align*}
a_{k}&\!=\!\ln\left|{\bf I}+\left(\sigma^2{\bf I}+\sum_{i\neq k}\bar{\bf \Gamma}_{ki}\right)^{-1}\bar{\bf \Gamma}_{kk}\right|
\!
\\&-\!
\text{{Tr}}\left(\!\!\!
\left(\!\!\sigma^2{\bf I}\!+\!\!\sum_{i\neq k}\bar{\bf \Gamma}_{ki}\!\!\right)^{-1}\bar{\bf \Gamma}_{kk}\!\!
\right)\!
-\!\frac{Q^{-1}(\epsilon_p)}{2\sqrt{n}}\!\!
\left(\!\!\sqrt{v_k}\!+\!\frac{2I}{\sqrt{v_k}}\!
\right)\!\!,
\\
a_{ck}&=\!\ln\left|{\bf I}\!+\!\left(\sigma^2{\bf I}+\sum_{i}\bar{\bf \Gamma}_{ki}\right)^{-1}\bar{\bf \Gamma}_{kc}\right|
\!
\\&-\!\text{{Tr}}\!\left(\!\!\!
\left(\!\!\sigma^2{\bf I}\!+\!\sum_{i}\bar{\bf \Gamma}_{ki}\!\!\right)^{-1}\!\!\bar{\bf \Gamma}_{kc}\!\!
\right)\!\!
-\!\frac{Q^{-1}(\epsilon_p)}{2\sqrt{n}}\!
\left(\!\!\sqrt{v_{ck}}\!+\!\frac{2I}{\sqrt{v_{ck}}}\!
\right)\!,\!\!
\\
{\bf A}_{k}&=\left(\sigma^2{\bf I}+\sum_{i\neq k}\bar{\bf \Gamma}_{ki}\right)^{-1}\bar{\mathbf{H}}_{k}\bar{{\bf W}}_{k},
\\
{\bf A}_{ck}&\!=\left(\sigma^2{\bf I}+\sum_{i}\bar{\bf \Gamma}_{ki}\right)^{-1}\bar{\mathbf{H}}_{k}\bar{{\bf W}}_c,
\\
{\bf A}_{ki}\!&\!=
\frac{Q^{-1}(\epsilon_p)}{\sqrt{nv_k}}
\left(\sigma^2{\bf I}+\sum_{n}\bar{\bf \Gamma}_{kn}\right)^{-1}\bar{\mathbf{H}}_{k}\bar{{\bf W}}_{i},
\\
{\bf A}_{cki}\!&\!=\frac{Q^{-1}(\epsilon_c)}{\sqrt{nv_{ck}}}\!\!
\left(\!\!\sigma^2{\bf I}\!+\!\bar{\bf \Gamma}_{ck}+\!\! \sum_{n}\bar{\bf \Gamma}_{kn}\!\!\right)^{-1}\bar{\mathbf{H}}_{k}\bar{{\bf W}}_i,
\\
{\bf B}_{k}&\!=\left(\sigma^2{\bf I}+\sum_{i\neq k}\bar{\bf \Gamma}_{ki}\right)^{-1}-\left(\sigma^2{\bf I}+\sum_{i}\bar{\bf \Gamma}_{ki}\right)^{-1}
\nonumber\\&
+\!
\frac{Q^{-1}(\epsilon_p)}{\sqrt{nv_k}}
\left(\sigma^2{\bf I}+\sum_{i}\bar{\bf \Gamma}_{ki}\right)^{-1}
\left(\sigma^2{\bf I}+\sum_{i\neq k}\bar{\bf \Gamma}_{ki}\right)
\\&\times
\left(\sigma^2{\bf I}+\sum_{i}\bar{\bf \Gamma}_{ki}\right)^{-1},
\\
{\bf B}_{ck}&\!=\left(\sigma^2{\bf I}+\sum_{i}\bar{\bf \Gamma}_{ki}\right)^{-1}-\left(\sigma^2{\bf I}+\bar{\bf \Gamma}_{kc}+ \sum_{i}\bar{\bf \Gamma}_{ki}\right)^{-1}
\nonumber\\&
+\!
\frac{Q^{-1}(\epsilon_p)}{\sqrt{nv_{ck}}}\!\!
\left(\!\!\sigma^2{\bf I}+\bar{\bf \Gamma}_{kc}+ \sum_{i}\bar{\bf \Gamma}_{ki}\!\!\right)^{-1}\!\!
\left(\!\!\sigma^2{\bf I}\!+\! \sum_{i}\bar{\bf \Gamma}_{ki}\!\!\right)
\\ &\times
\left(\!\!\sigma^2{\bf I}+\bar{\bf \Gamma}_{kc}+ \sum_{i}\bar{\bf \Gamma}_{ki}\!\!\right)^{-1}\!\!,\!\!
\end{align*}
where $I = \min(N_{BS}, N_u)$, $\bar{\bf \Gamma}_{ki}=\bar{\bf H}_k\bar{\bf W}_i\bar{\bf W}^H_i\bar{\bf H}_k^H $, $\bar{\bf \Gamma}_{kc}=\bar{\bf H}_k\bar{\bf W}_c\bar{\bf W}^H_c\bar{\bf H}_k^H $,  $\bar{\bf W}_c={\bf W}^{(j)}_c$, $\bar{\bf \Gamma}_{ki}=\bar{\bf H}_k\bar{\bf W}_i\bar{\bf W}^H_i\bar{\bf H}_k^H $, $v_{k}=2\text{Tr}\left(\left(\sigma^2{\bf I}+\sum_{i}\bar{\bf \Gamma}_{ki}\right)^{-1}\bar{\bf \Gamma}_{kk}\right)$,
$v_{ck}=2\text{Tr}\left(\left(\sigma^2{\bf I}+\bar{\bf \Gamma}_{kc}+\sum_{i}\bar{\bf \Gamma}_{ki}\right)^{-1}\bar{\bf \Gamma}_{kc}\right)$, and $\bar{\bf W}_k={\bf W}_k^{(j)}$ for all $i, k$.  
\end{lemma} 
Upon using \cite[Lemma 2]{soleymani2025framework} and Lemma \ref{lem:1}, we have
\begin{subequations}\label{opt-w-sur}
    \begin{align}    
	\max\limits_{{\bf W}_c,\{{\bf W}_k\}_{\forall k}, {\bf q}, {\bf z}} & \, \min_{k }\left\{ 2\eta_k^{(j)}z_k-\eta_k^{(j)^2}{p_k({\bf W}_c,{\bf W}_k)} \right\}\\
	\textrm{s.t.}\hspace{.5cm} & \tilde{r}_{pk}+q_k\geq \bar{r}_k^{th},
    \\
    & \sum_kq_k\geq \min_k(\tilde{r}_{kc}),
    \\ &\tilde{r}_{pk}+q_k-z_k^2\geq 0,
    \\&\eqref{eq:con2},
\end{align}
\end{subequations}
where ${\bf z}=\{z_1,z_2,\cdots,z_k\}$ is the set of auxiliary variables, and $\eta_k^{(j)}=\frac{\sqrt{r_k({\bf W}_c^{(j)},\{{\bf W}_k^{(j)}\}_{\forall k})}}{p_k({\bf W}_c^{(j)},{\bf W}_k^{(j)})}$. The updated beamforming matrices ${\bf W}_c^{(j+1)}$ and ${\bf W}_k^{(j+1)}$ are calculated  for all $k$ by solving \eqref{opt-w-sur}.

\subsection{RIS Optimization}
This subsection presents our solution to update ${\bf \Theta}^r$ and ${\bf \Theta}^t$ when ${\bf W}_c$ and ${\bf W}_k$ are fixed at ${\bf W}_c^{(j+1)}$ and ${\bf W}_k^{(j+1)}$. In this case, the optimization problem in \eqref{opt} is simplified to
\begin{subequations}\label{opt-t}  
\begin{align}  
	\max\limits_{{\bf q}, {\bf \Theta}^r, {\bf \Theta}^t} & \, \min_{k } \frac{r_{pk}+q_k}{p_k({\bf W}_c^{(j+1)},{\bf W}_k^{(j+1)})}\\
	\textrm{s.t.}\hspace{.5cm} & \eqref{eq:con1},\eqref{eq:con4},\eqref{eq:con3}.
\end{align}
\end{subequations}
This problem is indeed a maximization of the minimum weighted rate, since the transmission power $p_k({\bf W}_c^{(j+1)},{\bf W}_k^{(j+1)})$ is fixed when optimizing the STAR-RIS coefficients. 
To solve \eqref{opt-t}, we invoke an approach based on majorization minimization. More specifically, we first derive concave lower bounds for $r_{kc}$ and $r_{kp}$ to transform \eqref{opt-t} to a convex problem. We state the lower bounds in the following lemma. 
\begin{lemma}[\!\!\cite{soleymani2024rate}]\label{lem:2}
    The following inequalities hold for all feasible ${\bf H}_k$ for all $k$
\begin{multline*}
r_{pk}\!\geq\! \hat{r}_{pk}^{(j)}\! =\! 2\mathfrak{R}\!\left\{\!\text{{Tr}}\!\left(\!
{\bf A}_{k}\bar{\bf W}_{k}^H
{\mathbf{H}}_{k}^H\!\right)\!\right\}
+\!2\!\sum_{i\neq k}\!\mathfrak{R}\!\left\{\!\text{{Tr}}\!\left(\!
{\bf A}_{ki}\bar{\bf W}_{i}^H
{\mathbf{H}}_{k}^H\!\right)\!\right\}
\\
+a_{k}-
\text{{Tr}}\left(
{\bf B}_{pk}\left(\sigma^2{\bf I}\!+\!\!\sum_i\!{\bf H}_{k}\bar{\bf W}_{i}\bar{\bf W}_{i}^H{\bf H}_{k}^H\right)
\right),
\end{multline*}
\begin{multline*}
r_{ck}\geq \hat{r}_{ck}^{(j)} = a_{ck}
\\
+2\mathfrak{R}\left\{\text{{Tr}}\left(
{\bf A}_{ck}\bar{\bf W}^H_c
{\mathbf{H}}_{k}^H\right)\right\}
+2\sum_{i}\mathfrak{R}\left\{\text{{Tr}}\left(
{\bf A}_{cki}\bar{\bf W}_{c}^H
{\mathbf{H}}_{k}^H\right)\right\}
\\
-
\text{{Tr}}\!\!\left(\!\!
{\bf B}_{ck}\!\!\left(\!\!\sigma^2{\bf I}\!+{\bf H}_{k}\bar{\bf W}_c\bar{\bf W}^H_c{\bf H}_{k}^H+\!\!\sum_i\!{\bf H}_{k}\bar{\bf W}_{i}\bar{\bf W}_{i}^H{\bf H}_{k}^H\!\!\right)\!\!
\right),\!\!
\end{multline*}
where the parameters are defined in the same way as in Lemma \ref{lem:1}, except for $\bar{\bf W}_c={\bf W}^{(j+1)}_c $ and $\bar{\bf W}_k={\bf W}^{(j+1)}_k $ for all $k$.
\end{lemma}

Upon leveraging Lemma \ref{lem:2}, we can transform \eqref{opt-t} into the following convex problem
\begin{subequations}\label{opt-t2}  
\begin{align}  
	\max\limits_{{\bf q}, {\bf \Theta}^r, {\bf \Theta}^t} & \, \min_{k } \frac{\hat{r}_{pk}+q_k}{p_k({\bf W}_c^{(j+1)},{\bf W}_k^{(j+1)})} 
    \\
	\textrm{s.t.}\hspace{.5cm} & \hat{r}_{pk}+q_k\geq \bar{r}_k^{th},
    \\
    & \sum_kq_k\geq \min_k(\hat{r}_{kc}),
    \\ 
    &\eqref{eq:cm},\,\,\text{and}\,\,\eqref{eq:con3}.
\end{align}
\end{subequations}
Since \eqref{opt-t2} is convex, it is solvable by numerical tools such as CVX. The solution of \eqref{opt-t2} gives updated STAR-RIS coefficients ${\bf \Theta}^{r^{(j+1)}}$ and ${\bf \Theta}^{t^{(j+1)}}$. The overall algorithm converges to a stationary point of \eqref{opt} \cite{soleymani2025framework}. 

\section{Numerical Results}
\label{sec:ni}

We carry out Monte Carlo simulations to characterize the proposed energy-efficient scheme. 
In the simulation setup, we consider line-of-sight (LOS) propagation for the channels $\mathbf{D}$ and $\mathbf{D}_k$ for all users $k$, resulting in Ricean small-scale fading modeled according to \cite[(55)]{pan2020multicell}, with a Ricean $K$-factor of $3$. By contrast, the direct channels $\mathbf{G}_k$ are assumed to experience non-line-of-sight (NLOS) conditions and are modeled with Rayleigh fading. Large-scale fading for all channel links is captured using the path-loss model from \cite[(59)]{soleymani2022improper}. Additional propagation parameters, including antenna gains, bandwidth, noise power spectral density, path-loss exponent, and reference path-loss at 1 meter, are adopted from \cite[Appendix G]{soleymani2025framework}. 

The simulation parameters are configured as follows: the number of antennas at both the base station and each user is $N_{\mathrm{BS}} = N_u = 2$; the number of users is $K = 4$; the codeword lengths are set to $n=n_c = n_k = 256$ bits; and the error probability thresholds are $\epsilon=2\epsilon_c = 2\epsilon_k = 10^{-5}$. In terms of spatial deployment, users are evenly divided between the reflection and refraction regions of the STAR-RIS, as illustrated in Fig.~\ref{fig:network-model}. Consequently, a conventional reflective RIS would be unable to support half of the users. Again to overcome this limitation, we adopt a mode-switching operation for the STAR-RIS.

\begin{figure}[t]
    \centering
    \includegraphics[width=.8\linewidth]{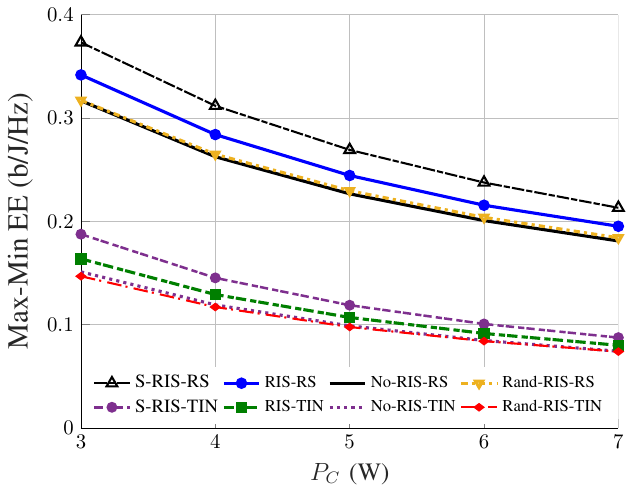}
    \caption{Max-min EE over $P_C$.}\vspace{-.5cm}
    \label{fig:sim1}
\end{figure}
Fig. \ref{fig:sim1} illustrates the relationship between the maximum minimum EE (max-min EE) and the static power consumption $P_C$ across different transmission schemes. The figure compares four variants of RSMA and treating interference as noise (TIN) schemes, each integrated with either a STAR-RIS, a reflective RIS, or no RIS, and also includes a baseline with randomly configured RIS elements. Note that the TIN/SDMA schemes are developed in \cite{soleymani2024optimization}. The results clearly demonstrate that RSMA schemes consistently outperform their TIN counterparts in terms of max-min EE across all values of $P_C$, affirming RSMA effectiveness in enhancing EE. Moreover, the incorporation of STAR-RIS provides additional but modest gains over conventional reflective RIS configurations, while--as expected--randomly chosen RIS parameters yield suboptimal performance. This comparison highlights the importance of STAR-RIS and the compelling synergy between RSMA and STAR-RIS in achieving robust, energy-efficient communication under FBL constraints.

\begin{figure}[t]
    \centering
    \includegraphics[width=.8\linewidth]{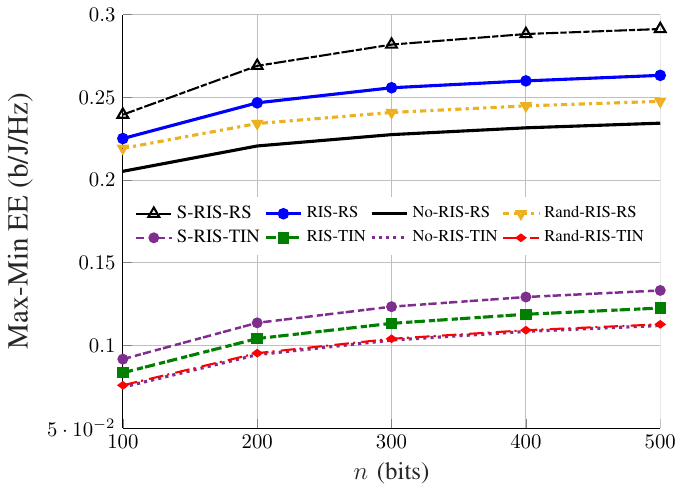}
    \caption{Max-min EE over $n$.}
    \label{fig:sim2}
\end{figure}
Fig. \ref{fig:sim2} demonstrates the average max-min EE as a function of $n$. The more stringent the latency requirement is, the shorter the codeword length becomes.  Fig. \ref{fig:sim2} reveals that the max-min EE attained reduces, when more restrictive latency requirement should be fulfilled.

\begin{figure}[t]
    \centering
    \includegraphics[width=.8\linewidth]{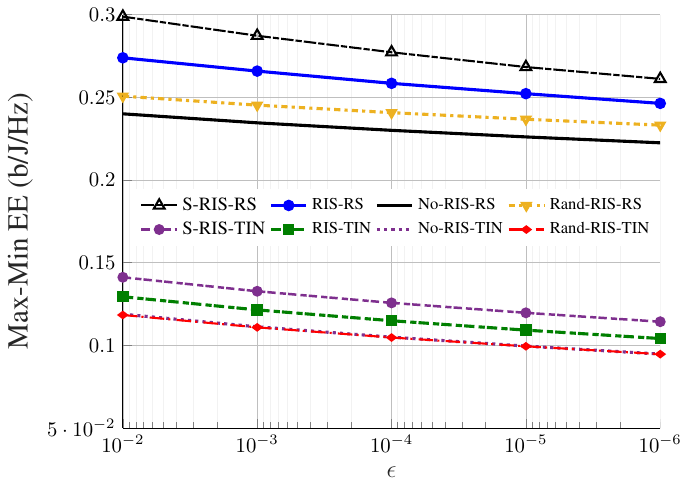}
    \caption{Max-min EE over $\epsilon$.}\vspace{-.15cm}
    \label{fig:sim3}
\end{figure}
Fig. \ref{fig:sim3} illustrates the variation in average max-min EE as a function of $\epsilon$, highlighting the impact of reliability. As shown in Fig. \ref{fig:sim3}, the max-min EE consistently decreases with smaller $\epsilon$, which correspond to more stringent reliability requirements. Each RSMA-based configuration outperforms its SDMA counterpart. Moreover, the use of STAR-RIS further improves EE, especially in RSMA schemes. 

\section{Conclusions}
Intrinsically integrating RSMA with STAR-RIS in MU-MIMO BCs significantly enhances EE in URLLC scenarios. By formulating and solving a max-min EE optimization problem using an AO and the FMP framework of \cite{soleymani2025framework}, the proposed energy-efficient scheme effectively addresses the challenges posed by stringent latency and reliability requirements. Numerical results confirm the superiority of the joint RSMA and STAR-RIS approach over conventional schemes upon varying system parameters such as the static power, blocklengths, and error tolerances, highlighting its potential as a key enabler for energy-efficient and robust communication in next-generation wireless networks.

\section*{Acknowledgment}
I. Santamaria’s work was funded by MCIN/AEI/10.13039/ 501100011033, under Grant PID2022-137099NBC43 (MADDIE) and FEDER UE, and by European Union’s (EU’s) Horizon Europe project 6G-SENSES under Grant 101139282. E. Jorswieck’s work was supported in part by the Federal Ministry of Education and Research (BMBF, Germany) through the Program of Souver\"an. Digital. Vernetzt. joint Project 6G-RIC, under Grant 16KISK031, and by European Union's (EU's) Horizon Europe project 6G-SENSES under Grant 101139282. 
Robert Schober’s work was funded by the German Ministry for Education and Research (BMBF) under the program of ``Souver\"an. Digital. Vernetzt.'' joint project 6G-RIC (Project-ID 16KISK023). L. Hanzo would like to acknowledge the financial support of the Engineering and Physical Sciences Research Council (EPSRC) projects under grant EP/Y026721/1, EP/W032635/1, EP/Y037243/1 and EP/X04047X/1 as well as of the European Research Council's Advanced Fellow Grant QuantCom (Grant No. 789028).

\bibliographystyle{ieeetr}
\bibliography{ref2.bib}

\end{document}